\documentclass[pra,preprint,showpacs,preprintnumbers,amsmath,amssymb]{revtex4}
\usepackage{epsf,graphicx}% Include figure files

\newcommand{\ket}[1]{\mathop{\left| #1 \right\rangle}\nolimits}
\newcommand{\bra}[1]{\mathop{\left\langle #1 \right|}\nolimits}
\newcommand{\braket}[2]{\mathop{\left\langle#1\right|\left.#2
\right\rangle}\nolimits}
\newcommand{\brawavy}[1]{\mathop{\left(#1\right|}\nolimits}
\newcommand{\ketwavy}[1]{\mathop{\left|#1\right)}\nolimits}
\newcommand{\braketwavy}[2]{\mathop{(#1\left.\right|#2
)}\nolimits}

\begin{document}

\title{Generalized quantum measurements. Part I: Information properties
of soft quantum measurements}

\author{Boris A.\ Grishanin}\email{grishan@comsim1.phys.msu.ru}
\author{Victor N. Zadkov}
\affiliation{International Laser Center and Faculty of Physics\\
M.\ V.\ Lomonosov Moscow State University, 119899 Moscow, Russia}

\date{June 6, 2005}

\begin{abstract}
A special class of soft quantum measurements as a physical model of the fuzzy
measurements widely used in physics is introduced and its information
properties are studied in detail.
\end{abstract}

\maketitle

\section{Introduction\label{section:intro}}

One of the fundamental transformations in quantum physics is the projective
measurement transformation, which sets a correspondence between quantum theory and
real physical experiment \cite{landauqm,mensky}. Despite the fact that physical
realization of this type of quantum transformations, which preserve the states of
the measurable quantum variables, i.e., \emph{nondemolition measurements}, causes
experimental difficulties, this field of experimental quantum physics is still of
prime importance \cite{Grangier1998,Pryde2004}.

From theoretical point of view, the projective measurement in its simplest
variant, i.e., with no addressing to its generalized variant of indirect
measurement, sets one-to-one correspondence between the orthogonal, i.e.,
entirely distinguishable, set of states $\ket k$ of the measurable system (we
will call it in the following simply the \emph{object}) and the set of
orthogonal states of the measuring device (we will call it in the following
simply the \emph{meter}) accompanied with a complete loss of the phase
relationships between them. In this case, the principal point is that one can
use for the measurement a \emph{classical} meter \cite{neumann}, which
corresponds to the case, when in the final state of the object--meter system
after the measuring process take part only those meter states for which their
quantum or microscopical nature is not essential.

Nowadays, however, progress in experimental quantum physics allows powerful
tools for preparation and manipulation with the quantum states \cite{QC}, so
that the standard concept of the quantum measurement as the projective
measurement transformation can be and has to be revised towards releasing the
limitation by the quasiclassical meter only and generalizing the concept of the
quantum measurement on the case when the meter is an essentially quantum
device. Such generalization of the quantum measurement concept is, obviously,
necessary for the adequate discussion of modern experiments in the field of
engineering of quantum information using such objects as atoms and ions in
various traps and photons of electromagnetic field. At the same time, the most
useful is not the most possible generalized concept of the quantum measurement,
as of an arbitrary quantum transformation, which contains information about the
measurable variables (see, for instance, \cite{vedral02,ozawa01, barnum02}),
but selection of those transformations, which, similar to projective
measurement, are based on one-to-one correspondence between the initial states
of the object and final states of the meter. The mechanisms of mapping the
quantum information via setting one-to-one correspondence between quantum
events, i.e., with the help of the corresponding classical information index
$k$, is undoubtedly the most important representation of quantum information
relations and has fundamental value for understanding of the basic grounds of
quantum mechanics.

In this work, we consider the so called soft quantum measurement, which is the
simplest model for a generalized quantum measurement that introduces in a
concentrated form all essential physical mechanisms responsible for inevitable
disturbance of the initial quantum information and the competitive character of
the process of its attainment, which results, as it is widely known, in the
possibility of reliable detection of the fact of its unauthorized usage.

The paper is organized as follows. In Sec.\ \ref{section:preserved}, the
general class of the information-preserving measurements is defined. In Sec.\
\ref{section:soft} the definition and general discussion of the soft
measurement is presented. In Sec.\ \ref{section:repeated} the general
properties of the repeated soft measurements are discussed. Sec.\
\ref{section:entangling} presents the quantitative analysis of the information
properties of the soft measurements. Finally, the key results of the paper are
summarized in conclusions.

\section{Quantum measurements preserving information\label{section:preserved}}

By definition, the mapping $\psi\to\varphi$ of the input states $\psi$ into the
output states $\varphi$ is represented by the projection operator $\ket\varphi
\bra\psi$ or a \emph{pointer}. For the measurement, we do understand under
$\psi$ only quantum states of the object, whereas $\varphi$ at the minimal
complete description must represent all essential under the measurement the
object--meter quantum states in the respected Hilbert space $H_A\otimes H_B$,
which, in the general case, must be supplemented with the reservoir $H_D$.

Obviously, the most direct realization of the classical measurement concept is the
transformation that is described with the set of projectors $\ket\varphi \bra\psi=
\ket{\ket k}\bra k$ with the classical index $k$ that numbers the input quantum
information. The content of the input information allows establishment of such
correspondence with the classical parameter due to the assumed orthogonality of the
input states $\ket k$ of the object. The latter form the orthogonal basis of the
space $H_A$  and are the eigen-states of a measurable physical variable $\hat A =
\sum\lambda_k \ket k\bra k$. For the measurement superoperator representation in the
form of the Kraus expansion $\sum_k\hat F_k^{} \odot\hat F_k^+$, where $\odot$ is
the substitution symbol for the transformed density matrix, it corresponds to the
operators $\hat F_k^{}=\ket{\ket k}\bra k$. The nondemolition projective measurement
corresponds to the choice $\ket{\ket k}=\ket k\ket k$, which realizes one-to-one
correspondence between the initial and the resulting bases of the measurable object,
which coincide, and those of the meter at the complete loss of the phase
relationships between their elements.

A simplest generalization of the projective measurement, which takes into
account essentially quantum character of the meter in the frame of the concept
of the ``ideal", i.e., nondemolition and absolutely precise, measurement is
given with the concept of the so called \emph{entangling measurement}
\cite{PRA2003,OptSpectr2004}. Following this concept, the entangling
measurement presents the measured information in the same form of
100\%-correlations, i.e., as one-to-one correspondence between the states of
the object and the meter but, in general case, does not destroy the coherency
of the measurable states $\ket k$ completely. Instead, this measurement
represents the resulted information in the form of quantum entanglement in the
bipartite object--meter system leaving the measurable states $\ket k$
unperturbed \cite{harocheRef}.

The entangling measurement maps the initial states $\psi=\sum c_k \ket k $ of the
object onto the resulted states $\psi_{ABD}^{}=\sum c_k\ket k\ket k \ketwavy k$ of
the tripartite system $A+B+D$ object--meter--reservoir, where $\ket k$, $\ket k$ are
the orthogonal bases in $H_A$ and $H_B$, respectively; $\ketwavy k$ is the set of
(not necessarily orthogonal) states of the reservoir with the scalar product
$R_{kl}= \braketwavy{k}{l}$, which is the so called \emph{entanglement matrix} that
represents the respective dephasing effects in the object--meter system
\cite{SPIE2004i,LLasPhys}. The corresponding mapping of the initial state of the
object onto the quantum states of the bipartite system object--meter are given by
the superoperator
\begin{equation}\label{Me}
{\cal M}_e=\sum\limits_{kl}R_{kl}\ket k\ket k\bra l\bra l\bra k\odot\ket l.
\end{equation}

\noindent Here the transformed density matrix $\hat\rho_A$ is determined in the
Hilbert space of the object $H_A$, whereas the transformation result
$\hat\rho_{AB}^{}$ is determined in the bipartite space $H_A\otimes H_B$. Eq.\
(\ref{Me}) does not include the initial state of the object because we assume
it either specified \emph{a priori} or an arbitrary one, but forgettable at the
process of measurement. This initial state does not affect the transfer of the
measurable information. At $R_{kl}=\delta_{kl}$, the considered superoperator
describes the standard projective measurement. At $R_{kl}\equiv1$, we have the
so called \emph{pre-measurement} \cite{schlosshauer}, which corresponds to the
total preservation of the initial coherency and transfer of all essentially
quantum (coherent) information, initially stored in the object, onto the set of
duplicated states $\ket{k}\ket k$ \cite{PRA00}.

In the latter case, we have exact cloning of the orthogonal set of the
measurable states and the superoperator (\ref{Me}) has the form ${\cal
M}_e=V\odot V^+$, where $V=\sum\ket k\ket k\bra k$. In terms of wave functions,
it corresponds to the isometric transformation $V$ from $H_A$ onto $H_A\otimes
H_B$. By contrast with the standard classical measurement, in the entangling
measurement the meter contains the exact value of the classical variable $k$,
but due to the entanglement in the object-meter system this information cannot
be red out and, specifically, be copied with the help of other physical systems
without respective losses of the coherent information, which is created during
the measurement. Without any loss, it is capable only in the case of the
standard projective measurement corresponding to a purely incoherent set of the
duplicated states $\ket{k}\ket k$.

Preserving the measured basis states in their initial form, which is the
essence of the nondemolition measurements, makes the entangling measurement a
very specific transformation realization of which (as well as the nondemolition
projective measurement) requires special efforts. From the information content
of the resulting state of the object--meter system, the nondemolition character
of the measurement gives no additional advantages, but defines how the coherent
quantum information in the output is linked with the initial set of the object
states $\ket{k}$.

One can generalize the entangling measurements in a natural way by giving up
the demand of nondestructiveness of the measurable states and, additionally,
the demand of one-to-one correspondence of the states of the system and the
meter. For this, in Eq.\ (\ref{Me}) the exactly cloned orthogonal states $\ket
k$ of the object and the meter can be replaced with the nonorthogonal states
$\ketwavy k$, which contain the internal indeterminacy and cannot be cloned, in
principle. After such a replacement the superoperator (\ref{Me}) remains
positively defined, but, as one can easily see, in order to preserve its
normalization it is necessary and sufficient to fulfill the condition
$R_{kl}Q_{kl}^AQ_{kl}^B=0$ for all $k,l$, where $Q_{kl}^{A,B}=\braketwavy
kl^{A,B}$ are the respective Gram matrix for the set of object and meter
states. This means that the orthogonality for the given $k,l$ must be hold true
at least in one of the subsystems of the object--meter--reservoir system, which
is due to the unitarity of the mapping, considered in the terms of the complete
system evolution. Respectively, for the completely coherent measurement,
$R_{kl}\equiv1$, the possibility of using nonorthogonal resulting states of the
object and the meter has an alternative character, i.e., for the output
object-system states we have either
\begin{equation}\label{I&II}
(I)\quad \ket{\ket k}=\ket k\ketwavy k\quad\mbox{or}\quad (II)\quad \ket{\ket
k}=\ketwavy k\ket k.
\end{equation}

The first considered possibility corresponds to the case of a nondemolition
``fuzzy" measurement, which confronts the distinct measurable states with not
entirely distinguishable states of the meter. The second possibility
corresponds to a particular case of an exact but ``destructive" measurement,
which changes the initial basis states of the object. Both these classes of
measurements are the measurements that completely \emph{preserve information}
in the joint states of the object--meter system about the measurable states of
the object regardless to the presence or absence of the external dephasing
(reservoir).

All the measurements considered in this paper, including those defined with the
superoperator (\ref{Me}), belong to the general class of \emph{nondemolition
measurements} that preserve the complete set of the classically compatible object
states $\ket k$, $\bigoplus_k\ket k=H_A$. The most general transformation for this
class of measurements is described by the superoperator of the form
\begin{equation}\label{Mgen}
{\cal M}_{\rm nd} = \sum_{kl} \bigl(\ket k\bra l \otimes \hat\rho^M_{kl}\bigr) \bra
k\odot\ket l,
\end{equation}

\noindent where the set of operators $\hat\rho^M_{kl}$ in Hilbert space $H_B$, which
describes essential for the measurement variables of the meter, defines the positive
block-type operator with the normalized diagonal terms, ${\rm Tr\,}\hat\rho^M_{kk}=
1$. This superoperator associates the object projectors $\ket k\bra l$ with the
$kl$-elements of the block-type operator $\bigl( \hat\rho^M_{kl}\bigr)=
\sum_{kl}\ket k\bra l\otimes\hat\rho^M_{kl}$ in the object--meter system, which, in
general case, describes the states of the meter entangled with the measurable states
of the object. Here, normalization of the diagonal terms ensures preserving of the
probability for the set of compatible measurable states of the object $\ket k\bra
k$, which are not perturbed during the measurement.

For the trivial case of $\hat\rho^M_{kl}=\hat\rho^B_0$, ${\cal M}_{\rm
nd}\hat\rho^A= \hat\rho^A\otimes\hat\rho^B_0$, all states of the meter are
associated with a single density matrix of the meter states, i.e., performing
no measurement. In case of the entangling measurement (\ref{Me}), we have a set
$\hat\rho^M_{kl}= R_{kl}\ket k\bra l$, which establishes correspondence
one-to-one between measurable states of the object with the similar, orthogonal
and, respectively, completely distinguishable states of the meter. Such a
measurement is the distinct one in the sense that for the compatible states of
the object and meter $\hat\rho^{AB}={\cal M}_{\rm nd}\hat\rho^A$ the joint
probability distribution for the respective events $\hat P^A_k=\ket k\ket
k\otimes\hat I_B$, $\hat P^B_l=\hat I_A\otimes\ket l\ket l$ is singular:
$P(k,l)={\rm Tr}\,\hat P^A_k\hat P^B_l\hat\rho^{AB} =\delta_{kl}\rho^A_{kk}$,
i.e. in the supporting subspace of the density matrix $\hat\rho^{AB}$ we have
$\hat P^A_k=\hat P^B_k$ for all $k$.

\section{Definition and physical essence of the soft measurement}\label{section:soft}

We consider here the most fundamental class of quantum nondemolition
measurements, which is characterized by using an \emph{``unclear''} set of
nonorthogonal states $\ketwavy k$ that contain the internal quantum uncertainty
for indication of the measurements results. This kind of measurement in the
limiting cases of the orthogonal or the trivial (consisted of the only state
$\ketwavy k\equiv\ket0$) sets is reduced to the described above entangling
(specifically, to projective) measurement and to the no measurement,
respectively. In the latter case, the only reason for the changes in the
object--meter system is the interaction of the object with the dephasing
subsystem, which takes place at $R_{kl}\neq1$. In a general case, the
measurements of this type, which just slightly change the initial state of the
object, are usually called the \emph{fuzzy} measurements
\cite{mensky,schlosshauer}. For this generalized measurement, the meter does
not contain any specified physical variable, which can store exact information
about the number $k$ of the measurable object states $\ket k$, and the attained
information is connected with the entire physical structure of the meter and is
represented in essentially quantum form. Note that the most comprehensive
description of the measurement is related to the consideration of the complete
object--meter system, whereas the quantum analysis of the measurement as the
transformation only in the space of the object states, which is performed in
many papers, does not reflect all essentially quantum information in the
object--system system.

We will call the measurement (\ref{Mgen}) the {\em soft} measurement, when the
resulting information of the meter is reflected in the matrix elements
$\hat\rho^M_{kl}=R_{kl}\ketwavy k\brawavy l$ by pure states $\ketwavy
k{\in}H_B$, uncertainty of which has purely quantum nature and connected with
their nonorthogonality, which leads to the impossibility of setting one-to-one
correspondence between the measurable orthogonal states similar to the
classically distinguishable  states of the meter. Such a measurement, described
with the superoperator
\begin{equation}\label{Mun}
{\cal M}=\sum\limits_{kl}R_{kl}\ket k\ketwavy k\bra k\odot \ket l\brawavy l\bra l,
\quad \braketwavy{k}{l}=Q_{kl},
\end{equation}

\noindent sets one-to-one correspondence between the object states $\ket k$ and
those of the meter, $\ketwavy k$. Its physical realizability is ensured by the
complete positivity \cite{kraus} of the transformation (\ref{Mun}) for
$R\geqq0$.

Physical essence of this transformation reduces to the independent from the
initial meter state transformation of the initial orthogonal basis states of
the object $\ket k$ into also orthogonal, i.e., completely distinguishable,
states $\ket k \ketwavy k$ of the bipartite object--meter system. At the same
time, the phase relationships between the initial states are generally
perturbed and their joint correlations are described with the matrix elements
$R_{kl}$, whereas the Gram matrix $Q$ describes the degree of quantum
distinguishability of the measuring states $\ketwavy k$ of the meter. The soft
character of the measurement is reflected by the difference of matrix $Q$ from
the identity matrix $Q=I$, which corresponds to the conventional (distinct)
entangling measurement. For the transformation (\ref{Mun}), as well as in the
case of distinct measurement, the classical content of the measurable object
states $\ket k$ is not perturbed, whereas the quantum information initially
stored in the initial object state $\hat\rho^A$ is redistributed between two
subsystems and perturbed due to dephasing.

In the limiting case of the trivial set $\ketwavy k\equiv\ket0$ the
superoperator (\ref{Mun}) corresponds to the independent from the meter
transformation of the object with partial loss of the joint coherency of the
measurable states $\ket k$. With this, the coherency of the object is preserved
in case of $R_{kl}=e^{i (\varphi_{k} - \varphi_{l})}$, i.e., with the
determinate phase transformation. \footnote{The given density matrix is
positively determined, which is necessary and sufficient for the complete
positiveness of the superoperator. Note that deviation from the phases
difference $\varphi_{k}$ in the exponent violates the positiveness.}

When the coherency is preserved, the soft measurement, considered as the
transformation in the bipartite object--meter system with the initially
``prepared" pure state $\ket0$ of the meter, is equivalent to the unitary
transformation. It maps the set of initial orthogonal states of the form $\ket
k\ket0$ onto the orthogonal states $\ketwavy k\ketwavy k$ and, obviously, can
be redefined up to the unitary operator $U_{AB}^{}$ in the total space
$H_A\otimes H_B$. The respective redefinition of the superoperator (\ref{Mun})
then can be represented with the superposition ${\cal U}{\cal S}$ of the
superoperator ${\cal S}=\ket0\bra0{\rm Tr}_B\odot$ of the resetting the meter
into the initial state $\ket0$ and of the unitary superoperator transformation
$\,{\cal U}=U_{AB}^{}\odot U_{AB}^{-1}$.

One can also easily see that the entropy of the initial object state is
entirely transferred into the entropy of the bipartite object--meter system,
$S[\hat\rho^A]= S[\hat\rho^{AB}]$. Respectively, the coherent information
\cite{schumacher}, defined with respect to its transformation from $H_A$ into
$H_A\otimes H_B$ \cite{PRA00}, is equal to its initial value $S[\hat\rho^A]$.
All the losses are due to only the dephasing and the respective violation of
the isometricity of the transformation at $R_{kl}\ne1$.

It is not difficult to calculate now the Hamiltonian of the transformation of
the infinitesimal fuzzy measurement in the object--meter system with the fixed
initial state $\ket0$ of the meter, which can be chosen as one of the resulting
states of the meter, i.e., $\ket0=\ketwavy0$. Calculating the infinitesimal
addition for a short time in the state of the object--meter system as the
result of the corresponding unitary transformation with the ``unperturbed'' or
``perturbation-free'' Hamiltonian $\hat\varepsilon=\sum_k\ket k\bra
k\otimes\hat\varepsilon_B^{}(k)$ and equating its result to the change, which
is caused by the transformation (\ref{Mun}), we do have $-i\frac{\Delta
t}\hbar\hat\varepsilon\ket k\ket0=\ket k\ketwavy{\delta k},$ where
$\ketwavy{\delta k}=\ketwavy k-\ket0$. From here, for the $k$-dependent
Hamiltonian of the meter  we receive $-i\frac{\Delta t}\hbar\bra
l\hat\varepsilon_B^{}(k)\ket0=\bra l\delta k)$, from which, due to the
hermicity, follows the equation for the uniquely determined matrix elements:
\begin{equation}\label{eps}
\hat\varepsilon_B^{}(k)=\sum_l\lim_{t\to0}i\frac{\hbar}{\Delta t}\bigl(\bra l\delta
k)\ket l\bra0-(\delta k\ket l\ket0\bra l\bigr),
\end{equation}

\noindent whereas other elements can be defined arbitrary way or, for example, be
set to zero. In the latter case, Eq.\ (\ref{eps}) has the structure $i(\hat a-\hat
a^+)$.

Generalization to the case when external dephasing is present can be described
with the unitary transformation in the system, which contains an additional
dephasing degree of freedom $H_D$, that, by analogy with the transformation
introduced in Ref.\ \onlinecite{LLasPhys} for the case of the entangling
measurement, includes also the infinitesimal unitary transformation created by
the Hamiltonian (\ref{eps}).

\section{Repeated measurements}\label{section:repeated}

\subsection{Measurements at the output of the meter}

The repeated application of the fuzzy measurement to the \emph{result} of the
initial measurement does not increase the attained information because the
resulted information contains an additional indeterminacy in comparison with a
single measurement, which does not vanish or decreases at the second (or
repeated) interaction of the meter with the object.

\subsection{Repeated measurements of the object with the accumulation of the
information \label{subsection:multy}}

For the repeated measurement of the \emph{object} with preservation of the
measurement results in independent degrees of freedom of the multicomponent
meter due to the $n$-fold application of the measurement transformation we
receive the following resulting transformation:
\begin{equation}\label{M^n}
{\cal M}^{(n)}=\sum\limits_{kl}R_{kl}^{\,n}\ketwavy k\dots\ketwavy k\ket k \bra
k\odot \ket l\bra l\brawavy l\dots\brawavy l.
\end{equation}

This transformation results in increasing incoherency and yields the multiply
duplicated unclear information about the value $k$. At the same time, the
quantum character of the measurable information is maximally preserved only in
the entire system (that includes all the meter's subsystems). After the
averaging over the $m<n$ output subsystems of the meter the remaining quantum
information is dephased and is characterized with the entanglement matrix
$R_{kl}^{(n-m)} =Q_{kl}^{\,m}R_{kl}^{\,n}$. This matrix defines the incoherency
of the measurement even without any dephasing during creation of the
entanglement in separate measurements, i.e., for $R_{kl}\equiv1$. In active
subspace $H_D={\rm sp}\{\ket{\ket k},\, k=1,\dots,D\}\sqsubseteq H^n_B$ of the
\emph{collective} states $\ket{\ket k}= \ketwavy k\dots\ketwavy k$ the
measurement transformation (\ref{M^n}) has the form of a single measurement,
but with entanglement matrices and scalar products corresponding to $n$
measurements.

Such measurement can be illustrated with many physical realizations. For
instance, if we select an atom from an atomic gas (ensemble) as a quantum
object, we can consider, in general, all the surrounding atoms as the
multipartite meter system. Then, separate collisions can be considered as the
separate measurement acts, which augments the measurable information in the
multipartite system. This situation, surely, exceeds the bounds of the standard
quantum measurement, which assures that all the measurable information is
accessible and can be used for any purpose.

\begin{figure}[ht]
\begin{center}
\epsfxsize=7cm\epsfclipon\leavevmode \epsffile{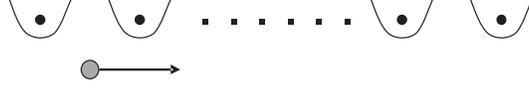}
\end{center}
\vspace{-0.02\textwidth}\caption{Physical realization of the repeated measurement
using the atomic optical dipole trap: an atom, a carrier of quantum information,
moves along the linear chain of atoms, which are located in the potential microtraps
of the optical dipole trap, each of which performs the measurement.\label{fig1}}
\end{figure}

Another physical situation with the atoms trapped in an optical dipole trap
\cite{granjier,meschede} fits our model of the repeated measurements more
precisely. In an optical dipole trap, an atom moving along the trapped in the
micropotential holes atoms performs repeated measurements (Fig.\ \ref{fig1}).
Successful experimental realizations of the nondemolition projective
measurements with a single photon \cite{haroche,Pryde2004} give us a hope that
the repeated measurements considered here would be realized experimentally in
the nearest future not only with atoms, but also with photons.

The dimension of the Hilbert space of the active states of the meter does not exceed
the dimension of the space with the entire set of states of the meter, despite the
fact that the space of the meter's states unrestrictedly expands. For appearance of
new active states different from the set $\ket k\ketwavy k\dots\ketwavy k$ (or to
the unitary equivalent to it) it is necessary that the dynamics along the different
degrees of freedom be independent and random. However, deviation of the bases used
in different measurements leads to the resulting states of the meter from the
indicated active space and if we have no \emph{a priori} information about these
deviations will result in losses of information about the object. Therefore, coding
of information in the series of repeated fuzzy measurements provides a resource for
the latent storing of information with the quantum key, which is unique.

When all the measurement results are preserved, the joint density matrix of the
object--meter system can be written as
\begin{equation}\label{rBnA}
\hat\rho^{\!B^n\!A}=\sum_{kl}R_{kl}^n\rho_{kl}^A\ketwavy k\dots\ketwavy k\ket k \bra
l\brawavy l\dots\brawavy l.
\end{equation}

\noindent In its turn, averaged over the states of the object density matrix of
the meter is
\begin{equation}\label{rBn}
\hat\rho^{\!B^n}=\sum\rho_{kk}^A\ketwavy k\dots\ketwavy k\brawavy k\dots\brawavy k,
\end{equation}

\noindent i.e., represented with the weighted sum of $D$ non-commuting
projectors. In orthogonal basis $\ket{\ket{e_k}}$ of the active subspace of the
collective states $H_D$, they can be rewritten in the form
\begin{eqnarray}
\hat\rho^{\!B^n\!A}&=&\sum\rho^{\!B^n\!A}_{ki,lj}\ket{\ket{e_k}}\ket i\bra
j\bra{\bra{e_l}}, \label{rBnAo} \\ \label{rBno}
\hat\rho^{\!B^n}&=&\sum\rho^{\!B^n}_{kl}\ket{\ket{e_k}}\bra{\bra{e_l}}
\end{eqnarray}

\noindent with matrix elements corresponding to the respective Eqs.\
(\ref{rBnA}) and (\ref{rBn}) and the choice of the basis $\ket{\ket{e_k}}$.

The matrix of scalar products for the vectors set $\ketwavy k\dots\ketwavy k$ has
the form $Q^{(n)}=\left(Q_{kl}^n\right)$. In the case of linearly independent set
$\ketwavy k$ for $n\to\infty$ this matrix has the form of identity matrix
$Q^{(\infty)}=I$. The orthonormalized basis in $H_D$ can be expressed via the
duplicated states of the meter with the help of the following relationship:
\begin{equation} \label{lc}
\ket{\ket{e_k}}=\sum_l\left({Q^{(n)}}^{-1/2}\right)_{kl}^* \ketwavy l \dots\ketwavy l.
\end{equation}

\noindent Here the corresponding formulas
\begin{eqnarray}
 \label{rhoBAkl} \rho^{\!B^nA}_{ki,lj}&=&\bra{\bra{e_k}}\bra i\hat\rho^{\!B^nA}
 \ket j \ket{\ket{e_l}}=
R_{ij}^n\,\rho^A_{ij}\,\psi_i^{B^n}(k)\psi_j^{B^n\!*}(l)\,,
\\ \label{rhoBkl} \rho^{\!B^n}_{kl}&=&\bra{\bra{e_k}}\hat\rho^{\!B^n}
\ket{\ket{e_l}}=\sum_{j}\rho^A_{jj}\,\psi_j^{B^n}(k)\psi_j^{B^n\!*}(l)
\end{eqnarray}

\noindent represent the states of the meter with the matrix
\begin{equation}\label{psinorm}
\psi_i^{B^n}(k)=\left({Q^{(n)}}^{1/2}\right)_{ki}
\end{equation}

\noindent of fixed dimension $D{\times}D$, the latter (to avoid confusion of
the ``dimension'' and ``matrix''!) does not depend on the number of
measurements $n$ and respective total dimension $D^n$ of the multipartite
Hilbert space $H^n_B$ of the meter. The entanglement matrix $R_{ij}$ is
essential only for constructing of the bipartite density matrix of the
object--meter system and does not affect the partial density matrix of the
meter because after the tracing over the object states their coherence is not
important. A set of non-orthogonal, in general case, functions
$\psi_i^{B^n}(k)$ makes sense of the ensemble of pure collective states of the
meter, post-selected after $n$ measurements and corresponding to the $i$-th
measured object states. All of them satisfy the normality condition
$\left(\psi_i^{B^n},\psi_i^{B^n}\right)\equiv1$, in which, due to definition
(\ref{psinorm}) of collective states, the scalar products reduce to the
normalized on unit the diagonal elements of the matrix $Q^{(n)}$.

\subsection{Two-level system}

Let us choose the representation in which $\ket0=(1,0)$, $\ket1=(0,1)$, and the
unclear set of measurable states has the form:
\begin{equation}\label{bas2D}
\ketwavy0=(1,0),\quad\ketwavy1=e^{i\chi}\left(\cos\frac\vartheta
2,\;e^{i\varphi}\sin\frac\vartheta 2\right).
\end{equation}

\noindent Then, matrices of the scalar products
\begin{displaymath}
Q=\left(\begin{array}{cc}
  1 & e^{i\chi}\cos\frac\vartheta 2 \\
  e^{-i\chi}\cos\frac\vartheta 2 & 1
\end{array}\right);\quad Q^{(n)}=\left(\begin{array}{cc}
  1 & e^{in\chi}\left(\cos\frac\vartheta 2\right)^n \\
  e^{-in\chi}\left(\cos\frac\vartheta 2\right)^n & 1
\end{array}\right),
\end{displaymath}

\noindent depend only on the angle $\vartheta$ between the vectors $\ketwavy
1$, $\ketwavy2$ and on its differential phase $\chi$.

Applying relations (\ref{eps}) to the two-level case we have
\begin{equation}\label{eps2D}
\hat\varepsilon_B^{}(0)=0,\quad\hat\varepsilon_B^{}(1)=\lim_{\Delta
t\to0}\frac\hbar{\Delta t}\left(\begin{array}{cc}
  -2\sin\chi\cos\frac\vartheta2 &ie^{i(\chi+\varphi)}\sin\frac\vartheta 2\\
  -ie^{-i(\chi+\varphi)}\sin\frac\vartheta 2 & 0
\end{array}\right)=\hbar\left(\begin{array}{cc}
-2\dot\chi&i\dot\vartheta\\
-i\dot\vartheta&0
\end{array}\right),
\end{equation}

\noindent where $\dot\chi$, $\dot\vartheta$ describe the rates of changes of
the respective angles in the process of the \emph{only} measurement. We do not
take into account here dependence on the second phase $\varphi$, which
describes the freedom in the choice of the common phase for the states
$\ketwavy0$, $\ketwavy1$.

For the matrix of wave functions of the measurable ensemble we have after $n$
measurements:
\begin{equation}\label{sqQ}
{Q^{(n)}}^{1/2}=\frac12 \left(\begin{array}{cc}
   \sqrt{1{-}\left(\cos\frac\vartheta 2\right)^n}
   {+}\sqrt{1{+}\left(\cos\frac\vartheta 2\right)^n} &
   e^{in\chi}\left[\sqrt{1{+}\left(\cos\frac\vartheta 2\right)^n}
   {-}\sqrt{1{-}\left(\cos\frac\vartheta 2\right)^n}\right]\\
   e^{-in\chi}\left[\sqrt{1{+}\left(\cos\frac\vartheta 2\right)^n}
   {-}\sqrt{1{-}\left(\cos\frac\vartheta 2\right)^n}\right]
   &\sqrt{1{-}\left(\cos\frac\vartheta 2\right)^n}
   {+}\sqrt{1{+}\left(\cos\frac\vartheta 2\right)^n}
\end{array}\right).
\end{equation}

\noindent Columns of this matrix play, in accordance with Eq.\ (\ref{psinorm}),
role of the wave functions describing $n$-fold excitations of the meter in the
minimal basis of the two-dimensional ($D=2$) space of the collective states.

Let us consider a sequence of $n$ identical measurements, performed with the
time period $T$ with small variation $\vartheta$ during the period time. We
will also assume that the relationships for the parameters necessary for the
asymptotically continuous changes of the result of the $n$-fold measurement as
a function of the continuous time $t$ at $n\propto t$. Respective continuous
dynamics has a quantum diffusion character, which shows at short times, plus to
the usual for the classical diffusion quadratic diffusional change of the
state, a specific linear diffusional change in the non-diagonal matrix
elements. Fluctuation character of the dynamics, which is typical for the
description of the classical diffusion process with the use of stochastic
equations, becomes apparent while considering the sequences of the respective
classically compatible variables in the space $H_B$, which are in charge for
the separate measurements in the whole measurement sequence. Their statistical
description cannot be reduced to a reversible dynamics of the collective
variables of the meter.

Matrix (\ref{sqQ}) at $\vartheta^2=4\varkappa T\to0$, $n=t/T\to \infty$,
$n\vartheta^2=4\varkappa t={\rm const}$, and $\chi=\dot\chi T$ has the finite limit
corresponding to the diffusion dynamics:
\begin{equation}\label{Qt}
Q^{1/2}(t)=\left(
\begin{array}{cc}
  \displaystyle s_+(t) & \displaystyle e^{i\dot\chi t}s_-(t)\\
  \displaystyle e^{-i\dot\chi t}s_-(t) & \displaystyle s_+(t)
\end{array}\right)\,,
\end{equation}

\noindent where $s_+(t)=\frac1 2 \bigl(\sqrt{1+e^{-\varkappa t}}+
\sqrt{1-e^{-\varkappa t}}\bigr)$, $s_-(t)=\frac12\bigl(\sqrt{1+ e^{-\varkappa
t}}-\sqrt{1-e^{-\varkappa t}}\bigr)$.

A similar limit has the entangling matrix
\begin{equation}\label{Rn}
 R^{(n)}=(R^n_{ij})=\left(\begin{array}{cc}
   1 & r^n \\
   r^{*n} & 1\
 \end{array}\right)\to\left(\begin{array}{cc}
   1 & e^{-\dot r t} \\
   e^{-\dot r^*t} & 1\
 \end{array}\right),\quad \dot r=\lim_{T\to0}\frac{1-r} T. \nonumber
\end{equation}

\noindent This matrix describes the dequantization of the measurement result
due to the external dephasing of the meter (reservoir). Simultaneously, at the
long time, asymptotic diagonalization of the collective states of the meter
occurs in accordance with the following asymptotic expression:
\begin{equation}\label{Qta}
Q^{1/2}(t) \stackrel{t\to\infty}{\Longrightarrow} \left(
\begin{array}{cc} \displaystyle1-\frac1 8\, e^{-2\varkappa t}&
\displaystyle\frac1 2\,e^{-\varkappa t+i\dot\chi t}\\
\displaystyle\frac1 2\,e^{-\varkappa t-i\dot\chi t} & \displaystyle 1-\frac18\,
e^{-2\varkappa t}
\end{array}\right). \nonumber
\end{equation}

\noindent this matrix describes how the soft measurement transforms into the
distinct completely coherent measurement (``pre-measurement") with the
orthogonal set of collective states of the meter $\ket{\ket k}\to\ket k$.

\subsubsection*{Partial density matrix of the meter}

From Eqs.\ (\ref{rBn}), (\ref{Qt}) we receive the partial density matrix of the
meter in the process of continuous measurement in the form:
\begin{equation}\label{rBc}
\hat\rho^{B^n}=\left(\begin{array}{cc}\frac12+\frac12\sqrt{1-e^{-\varkappa
t}}\left(\rho^A_{11}-\rho^A_{22}\right)&\frac1 2e^{-\varkappa t+i\dot\chi t}\\
\frac1 2e^{-\varkappa t-i\dot\chi t}&\frac1 2-\frac1 2\sqrt{1- e^{-\varkappa
t}}\left(\rho^A_{11}-\rho^A_{22}\right)\end{array}\right).
\end{equation}

\noindent At $t=0$, i.e., with no measurement, when for any initial states the
resulted state has the only one and the same value, $\ketwavy2=\ketwavy1
=\ket0$, the density matrix of the meter does not depend on $\hat\rho^A$ and is
equal to the projector $\ket0\bra0$ onto $\ket0$, which in the given basis
(\ref{lc}) for $n=1$ has the form $\ket0=(1/\sqrt2,1/\sqrt2)$. The choice of
the basis states of the meter, which determines in the structure of the
measurement superoperator (\ref{Mun}) how the measured information is
represented, is arbitrary.

At $t=\infty$, i.e., in the limit of infinitive series of the limiting
continuous soft measurements, the matrix (\ref{rBc}) becomes a diagonal one
with the matrix elements $\rho^A_{11}$, $\rho_{22}^A$, coinciding with the
respective matrix elements of the measurable object. This coincidence holds not
only for the limiting continuous measurements, but also for the series of
measurements with the finite accuracy about the object states $\ket k$, which
is determined with the tensor product of $n$ states $\ketwavy k\dots\ketwavy k$
in the form of multi-particle excitations of the $n$-fold copy of the
measurable systems, the information in the limit $n\to\infty$ is retrievable.

This leads to diagonalization of the marginal density matrix of the meter,
which is constructed of the duplicated states of the object--meter system $\ket
k \ketwavy k\dots\ketwavy k$, and to its coincidence with the density matrix of
the object. The described exact coincidence of the representations
$\hat\rho^{B^n} \to \hat\rho^A$ is due to the choice (\ref{lc}) of the basis
linked to the object variable $k$. The entropy of the quantum state of the
meter in the process of continuous measurement is changed from zero up to the
entropy of the object. At the same time, the absence of the internal
indeterminacy of the initial state of the meter is due to the absence of the
information about the structure of the meter, which is reflected in its partial
density matrix in the non-coherent, i.e., quasiclassical, form.

\subsubsection*{Joint density matrix of the object--meter system}

The set of the orthogonal states of the object--meter system $\ket k\ketwavy
k\dots\ketwavy k$ in the basis $\ket l\ket{\ket{e_m}}$ in accordance with Eqs.\
(\ref{rBnA}), (\ref{rhoBAkl}), and (\ref{Qt}), where $\psi^{B^nA}$ are
supplemented with the basis object states, has the form:
\begin{equation}\label{psiABi}
\psi^{B^nA}=\left\{\!\!\begin{array}{c}\displaystyle\vphantom{\int}\!\!\!
s_+(t) \ket{\ket{e_1}}\ket1+e^{i\dot\chi t} s_-(t) \ket{\ket{e_2}}\ket1\,,\\
\displaystyle\displaystyle e^{-i\dot\chi t} s_-(t) \ket{\ket{e_1}}\ket2+
s_+(t)\ket{\ket{e_2}}\ket2.
\end{array}\right.
\end{equation}

\noindent Using this set and in accordance with Eq.\ (\ref{rBnA}),
(\ref{rhoBAkl}) the density matrix of the object--meter system can be written
in the form of the $4\times4$-matrix
\begin{equation}\label{rBAt}
\hat\rho^{B^nA}(t)= \left(\begin{array}{cc}
  \rho_{11}^A\left(\begin{array}{cc}
  s_+^2 & s_+s_-e^{i\dot\chi t} \\
  s_+s_-e^{-i\dot\chi t}& s_-^2
  \end{array}\right) & \rho_{12}^Ae^{-\dot rt}\left(\begin{array}{cc}
                          s_+s_-e^{-i\dot\chi t} & s_+^2 \\
                         s_-^2e^{-2i\dot\chi t}& s_+s_-e^{-i\dot\chi t}
                         \end{array}\right)\\
   \rho_{21}^Ae^{-\dot r^*t}\left(\begin{array}{cc}
    s_+s_-e^{i\dot\chi t} & s_-^2e^{2i\dot\chi t} \\
    s_+^2 & s_+s_-e^{i\dot\chi t}
   \end{array}\right)
                       & \rho_{22}^A\left(\begin{array}{cc}
                         s_-^2 & s_+s_-e^{i\dot\chi t} \\
                         s_+s_-e^{-i\dot\chi t}& s_+^2
                         \end{array}\right)
\end{array}\right).
\end{equation}

\section{Information relationship in the processes of quantum measurements
\label{section:entangling}}

\subsection{Coherent information in the object--meter and object--object channels
\label{section:coherent}}

Key features of the coherent information \cite{schumacher} exchange at the
entangling measurement (\ref{Me}) are described in detail in Ref.\
\onlinecite{PRA2003}. The coherent information is purely quantum
\cite{Barnum01,SPIE2001a}) and, therefore, it cannot be copied or duplicated.
Thus, when the states $\ket k$ are duplicated, the coherent information is
transferred onto the superposition of bipartite states $\ket k\ket k$ and
lacking entirely in the channels ``initial--resulting state of the object" and
``initial state of the object--resulting state of the meter".

In accordance with its definition, the coherent information preserved in the
channel, which realizes the superoperator transformation ${\cal N}$, can be
written as
\begin{equation}\label{Ic}
I_c=S[{\cal N}\hat\rho]-S[\bigl({\cal N}\otimes{\cal I}\bigr)\Psi\Psi^+],
\end{equation}

\noindent where the first term describes the entropy at the output of the
channel and the second term---the so called exchange entropy that characterizes
the entropy surge due to the irreversibility of the transformation; the pure
state $\Psi$ describes the so called ``purified state" at the input of the
channel as the state of the bipartite system input+reference, which describes
the mixed input state as the result of its tracing over the auxiliary reference
system, and ${\cal I}$ is the identical transformation on the reference system
state, which is not perturbed.

Let us consider how the soft property of the measurement affects the
transformation of the coherent information. In the two-time channel
``object${\to}$object+meter", the soft measurement does not affect the coherent
information because of the orthogonality of the states $\ket k\ketwavy k$ of
the system object--meter. All losses of the coherent information in this
channel are due to the external dephasing only. By contrast, in the two-time
channel ``object--object" the influence of the soft measurement on the coherent
information is a nontrivial one---in this channel the amount of preserved
coherent information depends on both the external dephasing and the soft
character of the measurement. The latter determines how information is
distributed between the object and the meter.

Substituting in Eq.\ (\ref{Ic}) transformation ${\cal N}={\rm Tr}_B{\cal M}$
and taking into account Eq.\ (\ref{Mun}), we receive $I_c=S\bigl[ \bigl(R_{kl}
Q_{kl}\rho^A_{kl}\bigr)\bigr]- S\bigl[\bigl(R_{kl} Q_{kl}\Psi_{ki}^{}
\Psi_{lj}^*)\bigr]$, where in brackets ($\cdot$) are shown the matrix elements
$\tilde\rho_{kl}^{}$ and $\tilde\rho_{ki,lj}^{}$ corresponding to the
transformed density matrices of the object and object--reference system;
$\sum_i\Psi_{ki}^{} \Psi_{li}^*{=} \rho_{kl}^A$. For this channel, therefore,
contributions from the external dephasing and distinctness of the measurement,
which are presented with the respected matrices $R_{kl}$ and $Q_{kl}$, are
totally equivalent. Simplifying the expression in the argument of the second
term, one can rewrite the expression in the final form:
\begin{equation}\label{IcM}
I_c=S\bigl[\bigl(R_{kl}Q_{kl}\rho^A_{kl}\bigr)\bigr]-S\bigl[\Bigl(\sqrt{
\rho^A_{kk}}R_{kl} Q_{kl}\sqrt{\rho^A_{ll}}\Bigr)\bigr].
\end{equation}

In the absence of dephasing and at the maximal softness degree, i.e., $\ket
k\equiv\ket0$, the second term in Eq.\ (\ref{IcM}) vanishes, as far as its
argument comes to a density matrix of a pure state, whereas the first term
coincides with the entropy of the output, i.e., the coherent information is
transmitted without disturbance to the object system only. In the opposite
case, for the measurement with the complete distinguishability of the states of
the meter or for their maximal external dephasing $R_{kl}Q_{kl} = \delta_{kl}$,
both terms describe the entropy of the set of the measurable states of the
object $\ket k$, which is determined with the maximum entropy probability
distribution $p_k= \rho_{kk}^A$, and, respectively, the coherent information
vanishes due to the complete dequantization of the input information.

Calculation of the coherent information for a two-level system gives us the
following expression:
\begin{equation}\label{Icqmu}
I_c=\frac12[(1{-}x_1)\log_2(1{-}x_1){+}(1{+}x_1)\log_2(1{+}x_1)
{-}(1{-}x_2)\log_2(1{-}x_2){-}(1{+}x_2)\log_2(1{+}x_2)],
\end{equation}

\noindent where $x_{1,2}$ can be written with the only parameter $q=|R_{12}
Q_{12}|$ of the matrix $R_{kl}Q_{kl}$, diagonal matrix element $p=\rho_{11}^A$,
and the coefficient module of correlations $\mu=|\rho_{12}^A|/\sqrt{p(1-p)}$:
\begin{displaymath}
x_1=\sqrt{1- 4p(1-p)(1-q^2)},\quad x_2=\sqrt{1- 4p(1-p)(1-q^2 \mu^2)}].
\end{displaymath}

\begin{figure}[ht]
\begin{center}
\epsfxsize=0.45\textwidth\epsfclipon\leavevmode\epsffile{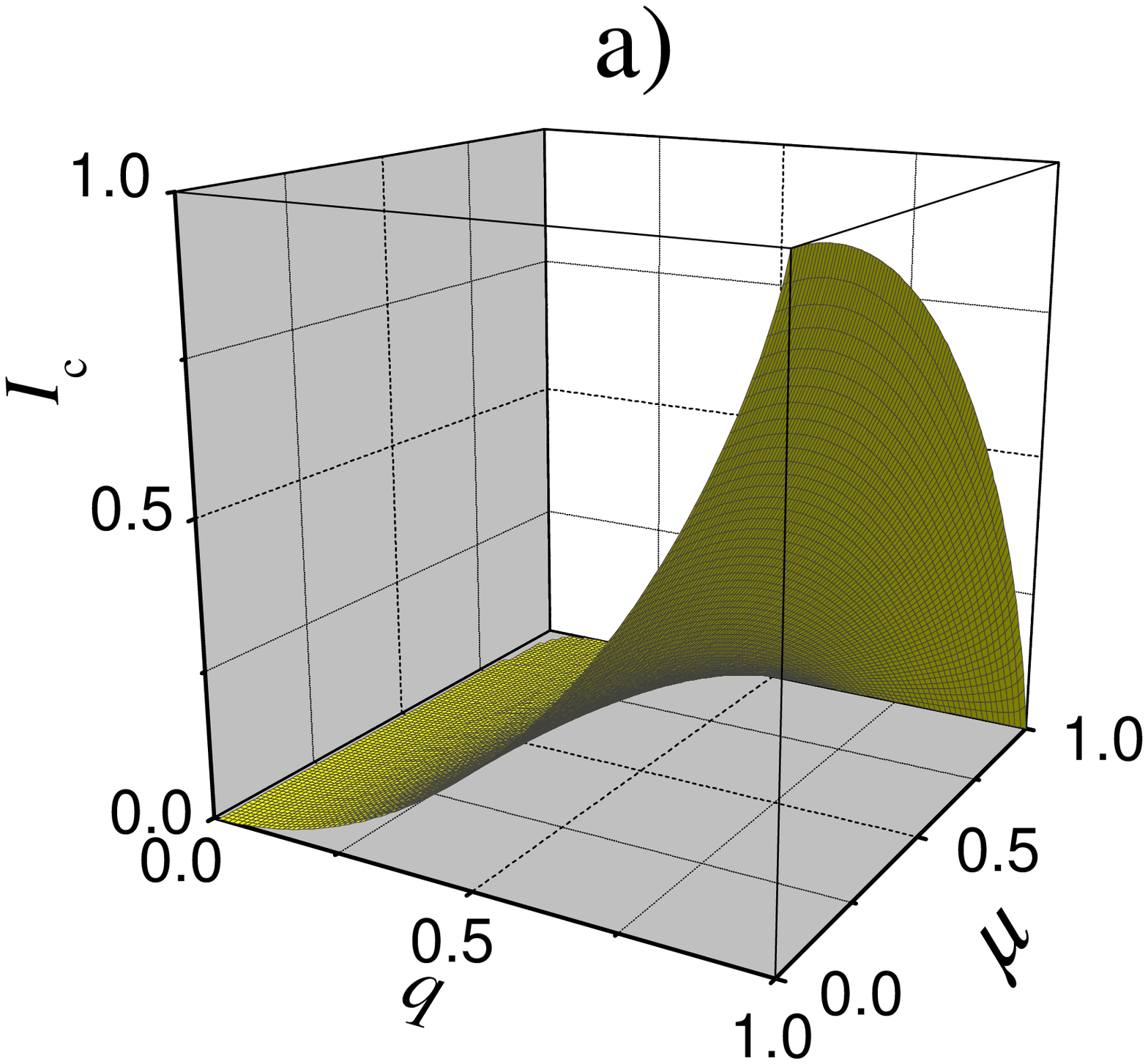}
\hspace{0.045\textwidth}
\epsfxsize=0.45\textwidth\epsfclipon\leavevmode\epsffile{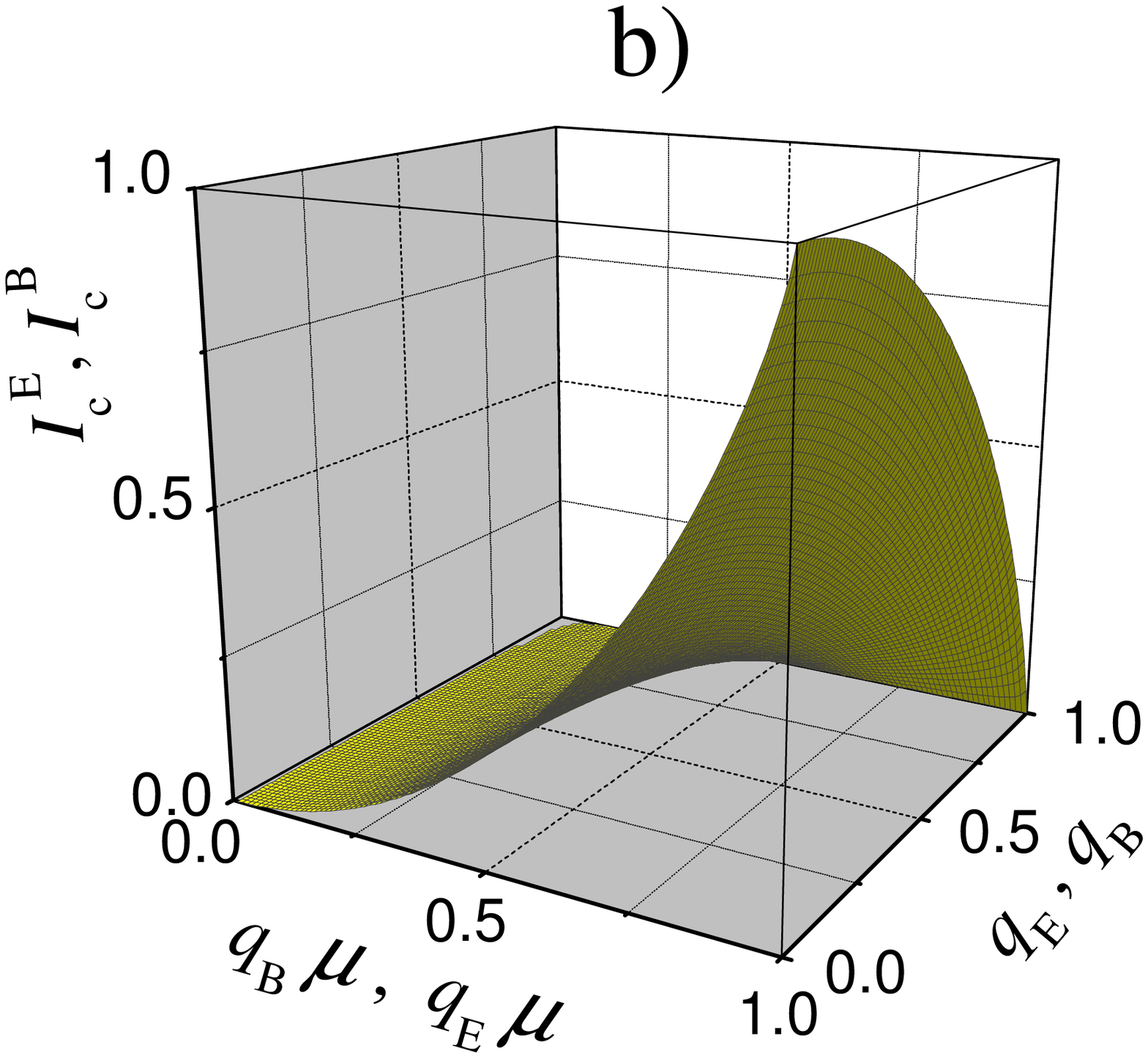}
\end{center}\vspace{-0.02\textwidth}
\caption{(a) The coherent information $I_c$, preserved in the object after the
soft measurement with the combined parameter $q$, which characterizes the level
of the softness and the degree of coherency of the measurement, versus the
function of the degree of coherency $\mu$ of the initial state of the object.
(b) The coherent information ${I_c^E}$ retrieved by Eve after the measurement
performed by Bob (curve 1) and the coherent information $I_c^B$ retrieved by
Bob after the measurement performed by Eve (curve 2) with the same fuzziness
parameter $q_B$. The graphs (a) and (b) numerically coincide, but they are
different by the physical content.} \label{fig2}
\end{figure}

\noindent The dependency corresponding to the Eq.\ (\ref{Icqmu}) for $p=1/2$
(for the maximally possible amount of information of the source equal to 1 bit)
is shown in Fig.\ \ref{fig2}a.

\subsection{Semiclassical information in the object--meter channel}

When one uses the object as a source of purely classical information in the
most general form of the mixed ensemble $\{p_\lambda,\hat\rho^A(\lambda)\}$ the
semiclassical information retrieved by the meter is described with the
respective ensemble $\{p_\lambda, \hat\rho_\lambda\}$ resulted after the
averaging over the object variables:
\begin{equation}\label{rBl}
\hat\rho_\lambda=\sum_k\rho^A_{kk}(\lambda)\ketwavy k\brawavy k.
\end{equation}

\noindent Non-classicality of this channel is related to its nonzero commutator
$\hat C_{\lambda\mu}=[\hat\rho_\lambda,\hat\rho_\mu]= \sum_k[\rho^A_{kk}
(\lambda)\rho^A_{ll}(\mu)-\rho^A_{kk}(\mu)\rho^A_{ll} (\lambda)]Q_{kl}\ketwavy
k\brawavy l$, which is nonzero only for the soft measurements with
$Q_{kl}\ne\delta_{kl}$. Nonorthogonality of the measurable states $\ketwavy k$
in ensemble (\ref{rBl}) leads to the respective reduction of the retrieved
information.

Let us illustrate how the amount of information depends on the parameters of
the soft measurement for the case of the multiple measurements in the same
basis for the input ensemble of pure states $\hat\rho_k^A=\ket k\bra k$. As one
can easily see, this ensemble corresponds to another one of the form  pure
states
\begin{equation}\label{rk}
\hat\rho^{B^n}_k=\ket{\ket k}\bra{\bra k} \nonumber
\end{equation}

\noindent in the active subspace of the meter, which is described in Sec.\
\ref{subsection:multy}. As an adequate quantitative characteristic for this
kind of the channels we can use the semiclassical information $I_s=S[\sum
p_k\hat\rho^{B^n}_k]-\sum p_kS[\hat\rho^{B^n}_k]$ \cite{Holevo}, which in this
case is simply equal to the entropy $S[\hat\rho^{B^n}]$ of the resulted density
matrix $\hat\rho^{B^n}=\sum p_k\ket{\ket k}\bra{\bra k}$, which have the matrix
elements (\ref{rhoBkl}) with $\rho^A_{kk}\to p_k$, at the output of the
measurement channel. For the two-level case, on account of Eq.\ (\ref{rBc}) and
for the a priori distribution $p_k=\{1/2,1/2\}$ we have the following amount of
information:
\begin{displaymath}
I_s=-\frac12\Bigl(\log_2\frac{1-e^{-4\varkappa t}}{4}+e^{-2\varkappa t}
\log_2\frac{1+e^{-2\varkappa t}}{1-e^{-2\varkappa t}}\Bigr),
\end{displaymath}

\noindent which monotonously changes from zero up to $I_{\max}=1$ with the
change of the measurement time $0\le\varkappa t\le\infty$ (in arbitrary units).
Effects of the external dephasing (reservoir) are not important in this case
because related phases of the measurable states of the meter are not essential.

\section{Competition at the retrieval of the information from the object
\label{section:concurr}}

\subsection{Competition at the retrieval of the coherent information
\label{susection:coherent}}

Let us consider now restrictions on the attained in the quantum measurement
information for the case when there is a single source of information (we will call
it ``Alice") and two receivers (``Bob" and ``Eve"), which retrieve this information
in series with the help of the repeated nondemolition quantum measurement with a
different choice of the measurable variables, in general case (by contrast with
Sec.\ \ref{subsection:multy}). The mapping of the quantum states in the respective
Hilbert spaces has the form $H_A\to H_A\otimes H_E\otimes H_B$ and the respective
complete superoperator transformation can be written as
\begin{equation}\label{AEB}
{\cal M}_{EB}=\bigl({\cal I}_E\otimes{\cal M}_B\bigr)\bigl({\cal
M}_E\otimes{\cal I}_B\bigr),
\end{equation}

\noindent where
\begin{displaymath}
\vphantom{\int_\sum}{\cal M}_E=\sum
R^E_{k_El_E}\ket{k_E}\ketwavy{k_E}\brawavy{l_E}\bra{l_E}\bra{k_E}\odot\ket{l_E}
\,, \quad {\cal M}_B=\sum R^B_{k_Bl_B}\ket{k_B}\ketwavy{k_B}\brawavy{l_B}
\bra{l_B}\bra {k_B}\odot\ket{l_B}
\end{displaymath}

\noindent describe the measurements performed by Bob and Eve under the same
object, but using different meters, and ${\cal I}_{E,B}$ is the respective
identical transformation over the variable of the meter inaccessible in this
measurement.

Expanding Eq.\ (\ref{AEB}), we have
\begin{displaymath}
{\cal M}_{EB}=\sum R^E_{k_El_E}R^B_{k_Bl_B}\braket{k_B}{k_E}\braket{l_E}{l_B}
\ket{k_B}\ketwavy{k_E}\ketwavy{k_B}\brawavy{l_B}\brawavy{l_E}\bra{l_B}\bra{k_E}
\odot\ket{l_E}.
\end{displaymath}

\noindent In the specific case of coinciding meter's bases , $\ket{k_B}=
\ket{k_E}$, we have
\begin{displaymath}
{\cal M}_{EB}^0=\sum R^E_{kl}R^B_{kl}\ket{k}\ketwavy {k} \ketwavy{k}
\brawavy{l}\brawavy{l} \bra{l}\bra{k}\odot\ket{l}.
\end{displaymath}

\noindent The difference between the considered above case and that one
described in Subsection\ \ref{subsection:multy} is in the independent use of
information contained here in bipartite states $\ketwavy k\ketwavy k$ of the
Eve--Bob system, which are coherently connected with the states $\ket k$ of the
object.

The result of the Eve's measurement does not depend on the subsequent Bob's
measurement only for the marginal states $\hat\rho^B$, but not for the joint
states $\hat\rho^{AE}$, which after the Bob's measurement (for instance in the
same basis) are represented instead of the initial superoperator ${\cal
M}_E=\sum R^E_{kl} \ket k\ketwavy k\brawavy l\bra l\bra k\odot\ket l$ with the
superoperator
\begin{displaymath}
{{\cal M}_E}'={\rm Tr}_B{\cal M}_{EB}^0= \sum R^E_{kl}R^B_{kl}Q^B_{kl}\ket
k\ketwavy k\brawavy l\bra l \bra k\odot\ket l,
\end{displaymath}

\noindent i.e. contains an additional dephasing factor $R^B_{kl}Q^B_{kl}$. An
absence of the back action, i.e., the equality ${{\cal M}_E}'={\cal M}_E$, in
the case of $R^E_{kl}\ne1$ when Eve retrieves the information in essentially
quantum form, is occurred only for the completely coherent measurement by Bob,
which contains no resulting information ($R^B_{kl}\equiv1$, $Q^B_{kl}
\equiv1$). In the case of completely dequantized measurement by Eve there is
never any reaction after the Bob's measurement, which ensures the stability of
classical information against its copying. A similar action does the
measurement by Eve on the Bob's measurement, which has after the Eve's
measurement the form ${{\cal M}_B}' ={\rm Tr}_E{\cal M}_{EB}^0=\sum
R^B_{kl}R^E_{kl}Q^E_{kl}\ket k\ketwavy k \brawavy l\bra l \bra k \odot \ket l,$
i.e., contains an additional factor $R^E_{kl} Q^E_{kl}$ comparing to the case
without Eve's measurement.

Such reaction of the quantum operation can be adequately described with the
respected reduction of the coherent information due to its reception by a new
receiver. In this case, due to the quantum measurement the meter receives the
coherent information about the object states only after the measurement and the
received information can be considered as the corresponding degree of quantum
entanglement in the object--meter system, which is measured, for instance, in
the system $A{+}B$ with the help of the difference $I_c=S[{\rm Tr}_B{\cal
M}\hat\rho_A]-S[{\cal M}\hat\rho_A]$, which is always positive in case of the
soft measurement.

By contrast with the similar definition used in ref.\ \onlinecite{PRA2003} for
the special case of a distinct entangling measurement, the first term here
determines the entropy of the object, but not the meter, because the entropies
of the object and the meter do not coincide for the case of the soft
measurement.

Calculating the respected information for Eve and Bob, we have
\begin{displaymath}
\begin{array}{c}\vphantom{\int_\int}I_c^E=S\bigl[\bigl(\rho^A_{kl}
R^E_{kl}
R^B_{kl}Q^E_{kl}Q^B_{kl}\bigr)\bigr]-S\bigl[\bigl(\rho^A_{kl}R^E_{kl}R^B_{kl}
Q^B_{kl}\bigr)\bigr],\\ I_c^B=S\bigl[\bigl(\rho^A_{kl} R^E_{kl}
R^B_{kl}Q^E_{kl}Q^B_{kl}\bigr)\bigr]-S\bigl[\bigl(\rho^A_{kl}R^B_{kl}R^E_{kl}
Q^E_{kl}\bigr)\bigr]\,.\end{array}
\end{displaymath}

\noindent The coherent information retrieved by Eve after Bob's measurement and
the amount of information received by Bob are shown for the two-dimensional
case in Fig.\ \ref{fig2}b as functions of the softness parameters for the Bob's
($q_B^{}=|Q^B_{12}|$) and the Eve's ($q_E^{}=|Q^E_{12}|$ ) measurements at
$R^B_{12}=R^E_{12}=1$ and for the density matrix of the object
$\rho_{11}^A=\rho_{22}^A=1/2$, $\rho_{12}^A= \rho_{21}^A=\mu/2$.

The respective analytical expressions for $I_c^E,I_c^B$ can be obtained using
Eq.\ (\ref{Icqmu}) for $I_c(q,\mu)$ that determines the coherent information
about the reference system, which is preserved in the object with the initial
density matrix $\hat\rho^A$ after the soft measurement with the softness
parameter $q$. This case corresponds to the change of variables $\{q\to
q_B^{}\mu,\,\mu\to q_E^{}\}$ when calculating information retrieved by Eve and,
respectively, $\{q\to q_E^{}\mu,\,\mu\to q_B^{}\}$ for the calculation of
information retrieved by Bob. In these calculations, $q_B^{}=1$ corresponds to
the case when Bob practically does not perform the measurement and retrieves
the unperturbed amount of Eve's information $I_c^E$ (and vice versa), which is
decreased with decreasing $q_B^{}$ due to the competence.

Note that the impact of the coherency parameter of the initial state $\mu$ on the
coherent information (\ref{Icqmu}), shown in Fig.\ \ref{fig2}a, and the competitive
information $I_c^{E,A}$, shown in Fig.\ \ref{fig2}b, is opposite. Whereas the value
of (\ref{Icqmu}) with increasing $\mu$ falls due to decreasing of the initial
entropy of the density matrix, which determines the entanglement between the object
and the reference system, the information $I_c^{E,A}$ increases with increasing
$\mu$ due to the respective increase of the object--meter entanglement after the
measurement. This entanglement does not exist for the incoherent mixture of pure
states $\{p_k,\ket k\}$ because of their imperturbability the states $\ket k$ even
at the completely coherent measurement are described with the incoherent mixture of
independent states $\ket k\ketwavy k$ of the object--meter system.

The competition character for the selection of the coherent information reveals
 in an opposite action of the parameters $q_B^{}$, $q_E^{}$ on the information
$I^E_c$ retrieved by Eve, for instance: with decreasing $q_E^{}$, i.e., with
increasing accuracy of the Eve's measurement, her information increases,
whereas with decreasing $q_B^{}$ it is also decreases up to the zero at
$q_B^{}=0$ due to the partial transfer of the information to Bob.

\subsection{Competition for selection of the classical information}

The competitive character of the quantum information is revealed also in the case of
semiclassical channels $A\to E$, $A\to B$ with the given ensemble of input states.
Ensemble corresponding to the first channel (\ref{rBl}), as it can be easily checked
by respective averaging of the superoperator (\ref{AEB}), is not modified after the
secondary measurement by Bob because he does not affects the input state of the
object. However, the measurement result by Bob depends on the basis choice, which is
used for the Eve's measurement. The respective interdependency of the resulting
quantum transformations lies in the basement of the quantum cryptography \cite{QC}.

Really, whereas the transformation ${\cal M}_{A\to E}{=}\sum_k\ketwavy{k}
\brawavy{k}\bra k\odot\ket k$ includes only parameters of the measurement
performed by Eve, namely, the measurable states $\ket k$ and the states of the
meter $\ketwavy k$, the transformation of the channel $A\to B$ depends also on
the parameters Eve's measurement, in accordance with (\ref{AEB}), as
\begin{equation}\label{MAtoB}
{\cal M}_{A\to B}{=}\sum_{kl}R^E_{kl}Q^E_{kl}\bigl({\cal P}_B\ket k_E\bra l_E\bigr)\bra
k_E\odot \ket l_E.
\end{equation}

\noindent Here ${\cal P}_B=\sum_k\ketwavy k_B\brawavy k_B\bra k_B\odot\ket k_B$ is
the superoperator of the \emph{[fuzzy] soft projection} from $H_A$ onto $H_B$, which
describes the result of the secondary measurement performed by Bob after the Eve's
measurement; $\ket{k,l}_{E,B}$ are the vectors of the measurable basis states in the
space $H_A$ of the object states for the measurements by Eve and Bob.

Introduced by Bob incoherency of the measurement due to the averaging over the
states of the object does not affect the retrieved by Bob information and the
soft character of the measurement is described with the superoperator ${\cal
P}_B$. At the same time, the incoherency of the Eve's measurement and its soft
character are reflected with the common matrix of dephasing (external in
respect to Bob) $q_{kl}^{}=R^E_{kl}Q^E_{kl}$, which describes the resulting
degree of the ``softness" of Eve's measurement. When the bases coincide, i.e.,
$\ket k_E=\ket k_B=\ket k$, the Eve's measurement does not affect the
information retrieved by Bob because ${\cal P}_B\ket k_E\bra l_E=
\delta_{kl}\ketwavy k_B\brawavy k_B$ and dependence on the parameters
$R^E_{kl},Q^E_{kl}$ is vanished. In this case, both Eve and Bob use only
classically compatible information about the object, which surely lacks the
specific quantum nature of the competition in its selection, and this
information can be copied independently. Dependency of the resulting channel
$A\to B$ on the Eve's transformations is related exclusively on the lack of
coincidence of their measurements bases, which makes essential quantum
disturbance of the object state introduced by Eve at $q_{kl}^{}\ne1$.

\begin{figure}[ht]
\begin{center}
\epsfxsize=0.45\textwidth\epsfclipon\leavevmode\epsffile{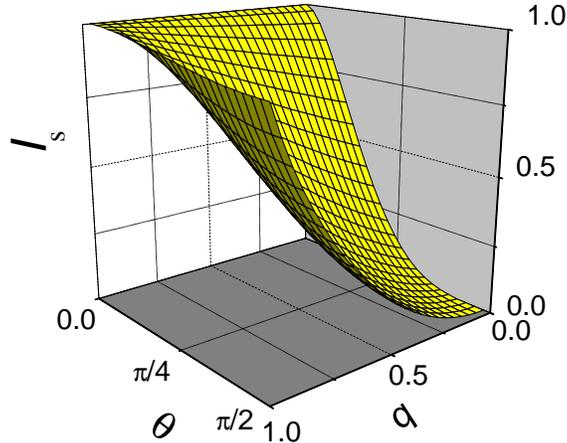}
\end{center}\vspace{-0.02\textwidth}
\caption{The semiclassical information $I_s$ versus the softness parameter
$q=|q_{12}^{}|$ of the measurement performed by Eve and the orientation angle
$\vartheta$ of the Eve's basis on the Bloch sphere in respect to the input
ensemble of two equiprobable pure states $\ket k_B$ at the rigid measurement by
Bob with $\ketwavy k_B=\ket k_B$.} \label{fig3}
\end{figure}

The corresponding generalization of the semiclassical channel (\ref{rBl}) on
account of its modification (\ref{MAtoB}) of the respective transformation of
its quantum input, has the form:
\begin{equation}\label{rBEl}
\hat\rho_\lambda=\sum_{kl}\rho^{A|E}_{kl}(\lambda)\,q_{kl}^{}\,{\cal P}_B\ket k_E \bra
l_E\,,
\end{equation}

\noindent where $\rho^{A|E}_{kl}$ is the density matrix of the object in the basis
Eve performs the measurement. The respected dependency of the semiclassical
information $I_s$ on the measurement parameters and on the input ensemble is shown
in Fig.\ \ref{fig3}. At its maximum degree the competition of the measurements
reveals at the orientation angle $\vartheta=\pi/2$, which, in the case of physical
realization of the Hilbert space of the object as the polarization degree of the
freedom of a photon, corresponds to the rotation of the linear polarization of the
photon at 45$^\circ$.

\section{Conclusions\label{section:conclusions}}

In conclusion, we introduced the model for selection of quantum information
with the help of the generalized quantum nondemolition measurement, which takes
into account the entanglement effects and, in most explicit form, summarizes
the fundamental differences between the quantum and classical information. In
particular, the possibility to consider the controlled degree of Eve's
interference while she uses the fuzzy measurement in order to attain the secret
information transmitted over the channel Alice--Bob, allows one using this
model of quantum measurement as the most simple, but conceptual enough model
for the interaction of the streams of quantum information in quantum
cryptography.

\acknowledgements

This work was partially supported by RFBR grant No.04--02--17554.


\begin{thebibliography}{99}
\bibitem{harocheRef}A typical example of physical realization of such a measurement
is given, e.g., by an experiment of \cite{haroche} on measurement of a number of
photons in a cavity with use of an atom passing through, if the measurement is
discussed on the detector response time scale, when the exact measurement result is
given by an intermediate joint atom--detector state, which coherence is only
partially destroyed.
\end{thebibliography}
\end{document}